\documentclass[a4paper,11pt]{article}

\usepackage{pos}
\usepackage{braket}

\newcommand{\one}{\, (\mathbf{1})}
\newcommand{\two}{\, (\mathbf{2})}
\newcommand{\otwo}{\, (\mathbf{12})}
\newcommand{\three}{\, (\mathbf{3})}

\newcommand{\othree}{\, (\mathbf{13})}
\newcommand{\tthree}{\, (\mathbf{23})}
\newcommand{\otthree}{\, (\mathbf{123})}

\newcommand{\RV}{\, (\mathbf{RV})}

\newcommand{\E}{{\mbox{\tiny{E}}}}
\newcommand{\eps}{\epsilon}

\newcommand{\al}{\alpha}
\newcommand{\nn}{\nonumber}
\newcommand{\T}{\mathbf{T}}
\newcommand{\hard}{\mathcal{H}}
\newcommand{\radState}{\{k_j,\lambda_j\}}

\newcommand{\s}{\mathcal{S}}
\newcommand{\J}{\mathcal{J}}
\newcommand{\beq}{\begin{eqnarray}}
\newcommand{\eeq}{\end{eqnarray}}
\newcommand{\npo}{{n+1}}
\newcommand{\npt}{{n+2}}
\newcommand{\npth}{{n+3}}
\newcommand{\pn}{\Phi_n}
\newcommand{\pnpo}{\Phi_\npo}
\newcommand{\pnpt}{\Phi_\npt}
\newcommand{\sig}{\sigma}
\newcommand{\LO}{{\mbox{\tiny{LO}}}}
\newcommand{\NLO}{{\mbox{\tiny{NLO}}}}
\newcommand{\NNLO}{{\mbox{\tiny{NNLO}}}}

\newcommand{\bbL}[2]{\overline{\bf L}^{#1}_{#2}}

\newcommand{\as}{\alpha_s}

\newcommand{\secn}[1]{Section~\ref{#1}}
\newcommand{\nnb}{\nonumber}

\newcommand{\varsi}{\varsigma}

\def\eq#1{eq.~(\ref{#1})}

\title{Refactorisation and Subtraction}

\author*[a]{Lorenzo Magnea}
\author[a]{Calum Milloy}
\author[b]{Chiara Signorile-Signorile}
\author[a]{Paolo Torrielli}

\affiliation[a]{Dipartimento di Fisica, Universit\`a di Torino, and INFN, 
Sezione di Torino \\ Via Pietro Giuria, 1 I-10125, Torino, Italy}

\affiliation[b]{Max-Planck-Institut f\"ur Physik, Boltzmannstrasse 8, 85748, 
Garching, Germany}

\vspace{3mm}

\emailAdd{lorenzo.magnea@unito.it}
\emailAdd{calumwilliam.milloy@unito.it}
\emailAdd{signoril@mpp.mpg.de}
\emailAdd{paolo.torrielli@unito.it}

\abstract{
Infrared subtraction algorithms beyond next-to-leading order necessitate the analysis  
of multiple infrared limits of scattering amplitudes, where several particles sequentially  
become soft or collinear. In this contribution, we report on the study performed in  
Ref.~\cite{Magnea:2024jqg}, which investigates these limits from the perspective of 
infrared factorisation, offering general definitions for strongly-ordered soft and 
collinear kernels, expressed in terms of gauge-invariant operator matrix elements. 
These definitions facilitate the identification of local subtraction counterterms 
for strongly-ordered configurations, whose integrals are designed to cancel the IR 
poles of real-virtual counterterms. This framework is validated at tree level for 
multiple emissions, and at one loop for single and double emissions.\\

\bigskip

MPP-2024-134
}

\FullConference{Loops and Legs in Quantum Field Theory (LL2024)\\
 14-19, April, 2024\\
Wittenberg, Germany\\}

\begin{document}
\maketitle


\section{Introduction}
\label{Intro}

High-order calculations of collider observables are crucial for achieving the theoretical 
precision required by current and upcoming experiments, and for identifying potential 
new physics signals. Such calculations require, in turn, complete control over infrared 
(IR) divergences, which becomes problematic at high orders in perturbation theory (see
Ref.~\cite{Agarwal:2021ais} for a recent review on the topic). At next-to-leading order 
(NLO), general and efficient algorithms for the subtraction of IR divergences have been 
developed~\cite{Frixione:1995ms,Catani:1996vz}, and are implemented in simulation 
codes heavily used by the experimental community. Extending subtraction algorithms to 
NNLO and beyond (see, for example,~\cite{Gehrmann-DeRidder:2005btv,Somogyi:2005xz,
Czakon:2010td,Caola:2017dug,Catani:2007vq,Boughezal:2015dva,Gaunt:2015pea,Magnea:2018hab})
is a challenging but necessary step to upgrade the accuracy standard of theoretical 
predictions, relevant to collider phenomenology\footnote{Recent developments in this field
are discussed in Ref.~\cite{TorresBobadilla:2020ekr} and in the references therein.}.
A significant obstacle in tackling this problem is the appearance, starting at NNLO, 
of \emph{strongly-ordered} singular configurations, which arise when two or more partons 
become unresolved at a different rate. There is a subtle interplay between these configurations
and the infrared poles of mixed real-virtual counterterms, which has not so far been
systematically understood. The present work is a contribution in this direction.

In what follows, we focus on the construction of strongly-ordered subtraction counterterms 
from the perspective of IR factorisation. We demonstrate how matrix elements of fields and 
Wilson lines, which describe the factorised emission of soft and collinear particles, can 
be `refactorised' in strongly-ordered configurations. This procedure provides formal 
expressions for strongly-ordered counterterms to all orders in perturbation theory, and 
identifies the pattern of cancellations between such counterterms and the singularities 
involving mixed real and virtual corrections. Our discussion is based on, but not limited 
to, the framework of \emph{local analytic sector subtraction}~\cite{Magnea:2018hab, 
Magnea:2018ebr,Magnea:2020trj,Bertolotti:2022aih,Bertolotti:2022ohq,Kardos:2024guf,
Chargeishvili:2024xuc}.


\section{The architecture of infrared subtraction}
\label{Archi}

To introduce our notation we start by considering the distribution of an IR-safe observable 
$X$, expanded in powers of the strong coupling as
\beq
\label{pertexpsig}
  \frac{d \sigma}{d X} \, = \, \frac{d \sigma_\LO}{d X} \, + \, 
  \frac{d \sigma_\NLO}{d X} \, + \, 
  \frac{d \sigma_\NNLO}{d X} \, + \, \ldots \, .
\eeq
The complexity of the subtraction problem emerges at NNLO, where the cancellation of IR 
divergences entails considering double-virtual corrections $VV_n$, integrated over an 
$n$-body phase space, together with real-virtual contributions $RV_{n+1}$, integrated 
over an $(n+1)$-body phase space, and with double-real radiation, $RR_{n+2}$, integrated 
in an $(n+2)$-body phase space. The observable NNLO distribution is, in fact, given by
\beq
\label{NNLOX}
  \frac{d \sig_\NNLO}{d X} = \lim_{d \to 4} 
  \Bigg[ \! \int d \pn \, VV_n \, \delta_n (X) + \int d \pnpo \, 
  RV_\npo \, \delta_\npo (X) 
  + \int d \pnpt \, RR_\npt \, \delta_\npt (X) 
  \Bigg] \, . 
\eeq 
The relevant squared matrix elements are
\beq
\label{defNNLO}
  VV_n \, = \,  \left| {\cal A}_n^{(1)} \right|^2 \! + 
  2 {\bf Re}  \left[ {\cal A}_n^{(0) \dagger} {\cal A}_n^{(2)} \right] ,
  \quad
  RV_\npo \, = \, 2 {\bf Re} \left[ {\cal A}_\npo^{(0) \dagger} \, 
  {\cal A}_\npo^{(1)} \right] , 
  \quad
  RR_\npt \, = \, \left| {\cal A}_\npt^{(0)} \right|^2 \! , \quad
\eeq
where $\mathcal{A}_{m}^{(k)}$ is the $k$-loop correction to the $m$-point scattering amplitude 
for the process under consideration. The double-virtual matrix element $VV_n$ features up to 
quadruple poles in the dimensional regulator $\eps=(4-d)/2$, while the real-virtual correction 
$RV_{n+1}$ displays up to double poles in $\eps$, and up to two phase-space singularities. 
Finally, the double-real matrix element $RR_{n+2}$ is finite in $d=4$, but it is affected by 
up to four phase-space singularities. At this order, we need to consider all possible 
double- and single-unresolved limits of $RR_\npt$, and carefully account for their overlap;
moreover, we have to properly subtract both the explicit poles and the single-unresolved 
limits of the real-virtual contribution. We collectively characterise single unresolved limits 
by the action of an operator $\bbL{\one}{}$ on squared real-emission matrix elements, 
defined in such a way as to remove any double counting. Similarly, double-unresolved limits 
are given by an operator $\bbL{\two}{}$, and their overlap by $\bbL{\otwo}{}$. The action of 
these operators on double-real and real-virtual contributions defines the minimal set of 
counterterms required for an NNLO subtraction approach. They read
\beq
\label{NNLOcountRR}
  & K^{\one}_\npt \, = \, \bbL{\one}{} \, RR_\npt \, ,  \qquad \quad 
  K^{\two}_\npt \, = \, \bbL{\two}{} \, RR_\npt \, ,  
  \\
  &
  K^{\otwo}_\npt \, = \, \bbL{\otwo}{} \, RR_\npt \, ,   \qquad \quad 
  K^{\RV}_\npo \, = \, \bbL{\one}{} \, RV_\npo \, .
\eeq
Upon designing a set of suitable phase-space mappings, to factorise resolved and unresolved  
phase-space measures, we can proceed to define {\it integrated counterterms}, as
\beq
\label{intcountNNLO}
  &
  I_\npo^{\, ({\bf 1})} \, = \, \int d \Phi_{\rm rad, 1}^\npt \, K_\npt^{\one} \, , \qquad
  I_n^{\, ({\bf 2})} \, = \, \int d \Phi_{\rm rad, 2}^\npt \, K_\npt^{\two} \, , 
  \\
  &
  I_\npo^{\otwo} \, = \, \int d \Phi_{\rm rad, 1}^\npt \, K_\npt^{\otwo} \, ,
  \qquad 
  I_n^{\RV} \, = \, \int d \Phi_{\rm rad, 1}^\npo \, K_\npo^{\RV} \, , 
\eeq
where the radiative phase spaces are defined by
\beq
\label{phaspafac2}
  d \Phi_{\npt} = \frac{\varsi_\npt}{\varsi_\npo} \, d \Phi_\npo \, 
  d \Phi_{{\rm rad}, 1}^\npt \, , \quad
  d \Phi_{\npt} = \frac{\varsi_\npt}{\varsi_n} \, d \Phi_n \, 
  d \Phi_{{\rm rad}, 2}^\npt \, , \quad  
  d \Phi_{\npo} \, \equiv \frac{\varsi_{\npo}}{\varsi_n} \, d \Phi_n \, 
  d \Phi_{{\rm rad}, 1}^\npo \, ,
\eeq
with $\varsi_p$ denoting the appropriate symmetry factors. Putting together the ingredients 
assembled so far, we can now write a fully subtracted form of the generic NNLO distribution,
\beq
\label{subtNNLO} 
  \frac{d \sig_\NNLO}{dX}
  & = &
  \int d \Phi_n \, \Big[ VV_n + I^{\two}_n + I^{\RV}_n \Big] \, \delta_n (X) 
  \\
  & + & 
  \int d \Phi_\npo \, \bigg[ \Big( RV_{\npo} + I^{\one}_\npo \Big) \, 
  \delta_\npo (X)
  \, -  \, \Big( K^{\RV}_\npo + I^{\otwo}_\npo \Big) \, \delta_n (X) \bigg]
  \nn \\
  & + & 
  \int d \Phi_\npt \, \bigg[ RR_\npt \, \delta_\npt (X) - 
  K^{\one}_\npt \, \delta_\npo (X) - \Big( K^{\two}_\npt - 
  K^{\otwo}_\npt \Big) \, \delta_n (X) 
  \bigg] \, . \nn
\eeq
We note that the third line is integrable in $\Phi_\npt$ by construction, since all singular 
regions have been subtracted with no double counting. In the second line, the integral
$I^{\one}_\npo$ cancels the $\epsilon$ poles of $RV_{\npo}$, but their combination is still 
affected by phase-space singularities. Those affecting $I^{\one}_\npo$ and $RV_{\npo}$ are 
cured by $I^{\otwo}_\npo$ and $K^{\RV}_\npo$, respectively: we conclude that the second line 
in \eq{subtNNLO} is free from phase-space singularities. On the other hand, there is in 
principle no guarantee that the $\epsilon$ poles of $K^{\RV}_\npo$ will cancel those of
$I^{\otwo}_\npo$, given the considerable degree of arbitrariness in constructing radiative 
counterterms. The goal of this note is to provide precise definitions of the counterterms that 
are geared towards making this cancellation automatic. A more detailed discussion is presented 
in Sec.~2.2 of~\cite{Magnea:2024jqg}. Having established the finiteness and integrability of 
both the second and the third line of \eq{subtNNLO}, the cancellation of poles in the first 
line directly follows from the KLN theorem. 

The arguments presented in this section can be formally generalised to N$^3$LO, and beyond. 
This requires organising a larger number of singular configurations and related overlaps. To
illustrate this, we can consider triple-real corrections arising at N$^3$LO: in that case, one 
needs to define unresolved limits and integrated counterterms as 
\beq
\label{N3LOcountRRR}
  K^{\, ({\bf h})}_\npth \, = \, \bbL{\, ({\bf h})}{} \, RRR_\npth \, , \quad \quad
  I_{n + 3 - q}^{\, ({\bf h})} \, = \, \int d \Phi_{{\rm rad}, q}^\npth \, K_\npth^{\, ({\bf h})} \, , 
  \quad \qquad 
{\bf h} \in \{ {\bf 3}, {\bf 13}, {\bf 23}, {\bf 123} \} \; , 
\eeq
where $q$ is the number of particles going unresolved at the highest rate. For instance, 
$\bbL{\three}{}$ collects the limits where three particles become unresolved at the same rate; 
$\bbL{\othree}{}$ corresponds to the strongly-ordered case where one particle becomes 
unresolved faster than the other two; $\bbL{\tthree}{}$ describes the configurations where two 
particles become unresolved at the same rate, but significantly faster than the third 
particle; finally, $\bbL{\otthree}{}$ describes the strongly-ordered scenario where each 
particle becomes unresolved at a different rate. Including mixed real-virtual corrections at 
one and two loops, one finds that N$^3$LO subtraction would require a total of 11 local 
counterterm functions, 5 of which involving strong ordering, with the remaining 6 
corresponding to uniform limits. Generalising further to N$^k$LO, the number of required 
counterterms turns out to be given by $c(k)=2^{k+1}-2-k$, of which only $k(k+1)/2$ 
corresponding to uniform limits: clearly, the problem of handling the cancellations between
integrated strongly-ordered and real-virtual counterterms becomes increasingly significant at 
higher perturbative orders.


\section{Democratic counterterms to any order}
\label{Demo}

In this section we exploit the factorisation properties of gauge amplitudes to find 
explicit expressions for the \emph{democratic} counterterms encoding uniform soft and 
collinear limits, following the discussion in Ref.~\cite{Magnea:2018ebr}. This approach relies 
on the knowledge of the IR structure of \emph{virtual} corrections, which is used as a starting 
point to infer suitable soft and collinear approximants of \emph{real} corrections. We refer 
to such a method as a \emph{top-down} approach, since we first analyse the first line of 
\eq{subtNNLO}, identify the required integrated counterterms, and then the corresponding 
integrands, which enter the second and the third lines.

The infrared factorisation formula for massless gauge-theory amplitudes 
reads~\cite{Dixon:2008gr,Gardi:2009qi,Becher:2009qa,Feige:2014wja} 
\beq
\label{AmpFact}
  \mathcal{A}_n \big( \{p_i\} \big) \, = \, \prod_{i = 1}^n \left[ 
  \frac{ \J_i \big( p_i, n_i \big)}{\J_{\E_i} \big( \beta_i, n_i \big)} \right] \, 
  \s_n \big( \{\beta_i\} \big) \, \hard_n \big( \{p_i\}, \{n_i\} \big) \, .
\eeq
The soft, jet, and eikonal jet functions $\s_n$, $\J_i$ and $\J_{\E_i}$ appearing in 
\eq{AmpFact} have explicit definitions (see for example Ref.~\cite{Agarwal:2021ais}), in terms 
of operator matrix elements involving semi-infinite Wilson lines aligned with the 
external-particle velocities $\beta_i$, 
\beq
\label{WilsonLine}
  \Phi_{\beta_i} (\infty,0) \, \equiv \, P \exp \left\{ {\rm i} g_s \T^a \int_0^\infty dz \,
  \beta_i \cdot A_a (z) \right\} \, ,
\eeq
as well as auxiliary Wilson lines along the directions $n_i$ ($n_i^2 \neq 0$), and quantum fields.
In \eq{WilsonLine}, $P$ denotes path ordering, $A_a$ represents the gluon field, and $g_s$ is 
the strong coupling constant. In the subtraction context, $\s_n$, $\J_i$ and $\J_{\E_i}$ can be 
considered special cases of more general functions, which can be used to model soft and 
collinear real radiation. For instance, the \emph{eikonal form factor} 
\beq
\label{eikFF}
  \s_{n,f_1 \ldots f_m} \big( \{ \beta_i \}; \radState \big) \, \equiv \, 
  \braket{\radState | \, 
  T  \prod_{i=1}^n \Phi_{\beta_i}(\infty,0) |0} \, ,
\eeq
describes the radiation of $m$ soft particles of flavours $f_j$, momenta $k_j$ and spin 
polarisations $\lambda_j$ ($j = 0, \ldots, m$), from $n$ Wilson lines representing hard 
particles, including virtual corrections in the soft approximation. Analogously, collinear 
radiation from an external particle $i$ can be modelled via \emph{collinear form factors}
such as
\beq
\label{collFFq}
  \mathcal{J}_{q, f_1 \ldots f_m}^{\alpha} \big(x; n; \{k_j, \lambda_j\} \big) \, \equiv \, 
  \braket{\radState | \, T \left[ \bar{\psi}^{\alpha}(x) \Phi_n(x,\infty) \right] |0} \, ,
\eeq
where we picked as an example a quark jet, $j = 1, \ldots, m$, and the case $m=1$ represents 
the purely virtual contribution, $\mathcal{J}_{f_i, f_i}=\mathcal{J}_{i}$. Similarly, the soft 
function in \eq{AmpFact} corresponds to the $m = 0$ case of \eq{eikFF}. Eikonal jets are 
obtained by replacing the quark field in \eq{collFFq} with an appropriate Wilson line, aligned 
with the classical quark trajectory.

At cross-section level, eikonal and collinear form factors must be squared, 
building up radiative soft and jet functions, which are fully local in the degrees of freedom 
of (multiple) soft and collinear real radiation. In the case of soft functions, the definition 
is straightforward~\cite{Magnea:2018ebr}, while collinear emissions require more care. Indeed, 
radiative jet functions must account for hard-collinear emissions to have non-zero momentum: 
therefore, at cross-section level, one of the two collinear form factors is evaluated at a 
shifted position $x$, which is Fourier-conjugate to the total momentum $\ell$ carried by 
final-state particles. We define then
\beq
\label{RadJetFuncs}
  J_{f, f_1 \ldots f_m}^{\al \beta} \big(\ell; n; \{k_j\}  \big) =
  \sum_{ \{ \lambda_j \} } \int d^d x \, {\rm e}^{ {\rm i} \ell \cdot x} \, 
  \mathcal{J}_{f, f_1 \ldots f_m}^{\al, \dagger} \big( 0; n; \{k_j, \lambda_j\} \big) 
  \mathcal{J}_{f, f_1 \ldots f_m}^{\beta} \big( x; n; \{k_j,\lambda_j\} \big) \, . 
\eeq
In~\eq{RadJetFuncs}, $f$ denotes the flavour of the parent particle, carrying the open spin 
indices $\al$ and $\beta$. Performing the $x$ integral will fix $\ell = \sum_j k_j$. At tree 
level, for $m=2$, it is straightforward to show that $J_{f, f_1 f_2}^{(0)}$ reproduces the 
tree-level splitting kernel for $f \rightarrow f_1+f_2$. Integrating~\eq{RadJetFuncs} over the 
radiative $m$-particle phase space, and summing over $m$, one finds
\beq
  \sum_{m = 1}^\infty \! \sum_{\{f_i\}} \int d \Phi_m \, 
  J_{q, f_1 \ldots f_m}^{\alpha \beta} = 
  {\rm Disc} \bigg\{ \! \int \! d^d x \, {\rm e}^{{\rm i} \ell \cdot x}
  \bra{0} T \Big[ \Phi_n (\infty, x) \psi^\beta (x) \bar{\psi}^\alpha (0) 
  \Phi_n (0, \infty) \Big] \! \ket{0}
  \!  \bigg\} \, .
\label{CompleteColl}
\eeq
The r.h.s.~of \eq{CompleteColl} is the discontinuity of a two-point function in the presence of 
Wilson lines, which can be shown to be IR finite order by order. Such a \emph{finiteness} 
condition is crucial for applying factorisation arguments to the construction of local IR 
counterterms at any order in perturbation theory \cite{Magnea:2018ebr}: similar conditions 
apply for soft functions and eikonal jets. Indeed, by expanding the l.h.s.~of \eq{CompleteColl},
and of its analogue for the soft function, we find order-by-order finiteness conditions
that embody the KLN cancellations. At NLO, for example
\beq
\label{FinCondNLO1}
  S_n^{(1)} \big( \{ \beta_i \} \big) + \int d \Phi(k) \, S_{n, g}^{(0)} \big( \{ \beta_i \}; k \big) 
  \! & = & \! \text{finite} \, , \\
\label{FinCondNLO3}  
  \sum_{f_1} \int d \Phi(k_1) J_{f, f_1}^{(1) \alpha \beta} (\ell; k_1) + \sum_{f_1, f_2} 
  \varsi_{f_1 f_2} \! \int d \Phi(k_1) d \Phi(k_2) \, 
  J_{f, f_1 f_2}^{(0) \, \alpha \beta} (\ell; k_1, k_2) \! & = & \! \text{finite} \, ,
\eeq
where $\varsi_{f_1 f_2}$ is a phase-space symmetry factor. The conditions in 
eqs.~(\ref{FinCondNLO1}-\ref{FinCondNLO3}) immediately suggest that the integrands of the
real-radiation contributions can serve as candidate soft and collinear local NLO counterterms. 
At NNLO, the analogues of eqs.~(\ref{FinCondNLO1}-\ref{FinCondNLO3}) involve a double-virtual 
correction, a real-virtual term and a double-radiative function, as expected from general 
IR-cancellation theorems. It is important to notice that counterterms identified via 
finiteness relations directly reproduce only uniform infrared limits. The construction of 
strongly-ordered counterterms from factorisation is discussed in the next section.


\section{Strongly-ordered counterterms to any order}
\label{Dicta}

In this section we aim to express strongly-ordered limits in terms of universal operator matrix
elements, in the spirit of factorisation. To deduce their form, we start by considering uniform 
double-unresolved limits, and applying strong-ordering conditions. We analyse the soft limit 
first. The tree-level double-soft current for gluons 1 and 2 with momenta $k_1$ and 
$k_2$~\cite{Catani:1999ss} simplifies considerably in the strongly-ordered limit in which 
$k_2 \ll k_1 \ll \mu$, with $\mu$ the hard scale of the process. The corresponding form factor 
is given by an interesting `refactorisation' of the double-radiative soft function:
\beq
\label{softrads.o.}
  \Big[{\cal S}^{(0)}_{n; \, g, g} \Big]^{a_1 a_2}_{\{d_i e_i\}}
  \big(\{\beta_i\};k_1, k_2) & \equiv &  
  \bra{k_2, a_2} \, 
  T \bigg[
  \Phi^{\,\, a_1 b}_{\beta_{k_1}} (0,\infty) 
  \prod_{i = 1}^{n} \Phi^{\quad \,\,\,\,\, c_i }_{\beta_{i}, \, d_i}(\infty,0) \bigg] \, \ket{0} \\
  & & \hspace{1cm}
  \times \,
  \bra{k_1, b} \,  
    T \bigg[
    \prod_{i=1}^{n}
  \Phi_{\beta_i, \, c_i e_i}(\infty,0)
  \bigg]  \ket{0} \Big|_{\rm tree} \nn \\
  & = & \left[ {\cal S}^{(0)}_{n+1, \, g} \right]^{a_2, \, a_1 b}_{\{d_i c_i\}}
  \left(\beta_{k_1}, \{\beta_{i}\}; k_2 \right) \, 
  \left[ {\cal S}^{(0)}_{n, \, g} \right]_{b, \, \{ c_i e_i \}} \! 
  \left(\{\beta_{i}\};k_1 \right)\, .
  \nn  
\eeq
Eq.~(\ref{softrads.o.}) can be interpreted as follows: gluon 1 is soft compared to the $n$ hard 
Born partons, but appears as hard when probed by gluon 2, with $k_2 \ll k_1$. The original 
system of $n$ Wilson lines thus radiates the harder gluon 1, which then `\emph{Wilsonises}': 
indeed, at this stage, the new system of $(n+1)$ Wilson lines radiates the softer gluon 2. This 
is described by a factorised matrix element, where gluon 2 remains a final-state parton, while 
gluon 1 plays the double role of final-state parton (when radiated by one of the $n$ original 
hard legs) and of Wilson line in the adjoint representation (when radiating gluon 2). 
Tree-level soft refactorisation has been tested against the expressions in
Ref.~\cite{Catani:2019nqv} up to  3 gluons, and it is natural to conjecture its validity for
any number of gluons.

In the case of multiple collinear emissions, the situation is more involved due to spin 
correlations, but a refactorised form can still be identified. For instance, the NNLO 
strongly-ordered collinear configuration for a $q \, \to \, q_1' \, \bar q'_2 \, q_3$ 
branching is given by
\beq
\label{FGs.o.}
  \lim_{\theta_{12} \ll \theta_{13} \to 0} RR_{\npt}
  & = & 
  \frac{(8 \pi \as)^{\, 2}}{s_{12} \, s_{[12]3}} \,
  \, P^{\rho\sigma}_{q \to gq} \big( z_{[12]}, q_\bot \big) \, 
  d_{\rho \mu} \big( k_{[12]} \big) \nn \\
  && \hspace{2cm} \times \, P^{\mu \nu}_{g \to q \bar q} 
  \left(z_1/z_{[12]}, k_\bot \right) \, 
  d_{\sigma\nu} \big( k_{[12]} \big) \, B_n \, , 
\eeq
where the intermediate-particle momentum is $k_{[12]} \equiv  k_1 + k_2$, its collinear 
energy fraction is $z_{[12]} \equiv z_1 + z_2 = 1 - z_3$, and $s_{[12]3} = 2 \, k_{[12]} 
\cdot k_3$. Finally, $d_{\mu \nu}(k) = - g_{\mu \nu} + (k_\mu n_\nu + k_\nu n_\mu)/(k 
\cdot n)$ is the intermediate gluon polarisation sum. As expected, the same result can be 
written in terms of an appropriate convolution of radiative jet functions:
\beq
\label{eq:coll_str_ord_example1}
  && \int \frac{d^d \ell}{(2\pi)^d} \,
  \left[ \lim_{\theta_{12} \ll \theta_{13} \to 0}
  J_{q,qq'\bar q'}^{(0)}(\ell;k_1,k_2,k_3) \right]
  \, \equiv \, 
  \int \frac{d^d \ell}{(2\pi)^d} \,\,
  J_{q,gq;g,q'\bar q'}^{(0)} (\ell;k_1,k_2,k_3) \quad \nn \\
  && \hspace{2cm} = \,  \int \frac{d^d \ell}{(2\pi)^d} \,\,
  J^{\, ; \rho\sigma \, (0)}_{q,gq}\big( \ell; k_{[12]}, k_3 \big) \, 
  \int \frac{d^d \ell'}{(2\pi)^d} \,\,
  J^{\rho\sigma \, (0)}_{g, q' \bar q'}(\ell';k_1,k_2) \, ,
\eeq
where the first line introduces the notation for a strongly-ordered jet function, outlining the 
sequential splittings involved. The integrals over $d^d \ell$ and $d^d \ell'$ address the 
delta-function constraints in the corresponding jet-function definitions, which ensure that the 
parent parton momentum equals the total momentum of its decay products. Spin indices of the 
parent quark are not explicitly shown, and are replaced by the semicolon. The Lorentz spin 
indices after the semicolon correspond to the daughter gluon produced by the splitting.
Analogous versions \eq{eq:coll_str_ord_example1} hold for splittings of different flavours, and
generalise to larger numbers of emitted particles~\cite{Magnea:2024jqg}.

As mentioned in \secn{Intro}, strongly-ordered counterterms have to combine, upon integration 
over the most unresolved parton, with real-virtual counterterms, cancelling their poles. In 
order to make such an interplay manifest, it is useful to exploit once more the idea of 
refactorisation. We need to consider one-loop radiative soft and jet functions: for the sake 
of illustration, we focus on the former. We note that, contrary to their virtual counterpart, they 
do not reduce to pure counterterms, and contain both IR poles and finite contributions. For 
our purposes, they can be considered as scattering amplitudes with Wilson-line sources, which 
points to a natural factorisation of their virtual IR poles. Indeed, for example, applying the 
standard soft-jet-hard factorisation to the single-radiative soft function leads to 
\beq
\label{eq:soft_fact_nlo}
  \s_{n,g} \big( \{\beta_i\}; k \big)
  \, = \,
  \s_{n+1} \big( \{\beta_i\} , \beta_k) \,
  \frac{\J^{\mu}_{g,g}(0;k)}{\J_{\E_g}(\beta_k)} \,
  \s_{n,g}^{\mathcal{H},\mu} \big( \{\beta_i\}; k \big) \, ,
\eeq
where the factor $\s_{n,g}^{\mathcal{H},\mu}$ is finite in $d=4$. Expanding to one-loop order, the terms containing IR poles are thus
\beq
\label{eq:Sng1}
  \s_{n,g}^{(1)} \big( \{\beta_i\}; k \big)
  \, = \, &
  \left[ \s_{n+1}^{(1)} \big( \{\beta_i\}, \beta_k \big) -
  \J_{\E_g}^{(1)} (\beta_k) \right] \s_{n,g}^{(0)} \big( \{\beta_i\}; k \big) 
  \nn \\ &
  + \, 
  \J^{(1) \mu}_{g,g} (0;k) \, \s_{n,g}^{(0) \mu} \big( \{\beta_i\}; k \big) \, .
\eeq
We will show in the next section how to reconstruct from \eq{eq:Sng1} (upon squaring) the soft 
contribution to $K^{\RV}_\npo$, plus hard-collinear corrections, which will need to be subtracted. 
The remaining soft poles will naturally cancel against those arising in the integrated strongly-ordered
soft counterterm.


\section{Engineering cancellations}
\label{Engi}

In this section we present a construction of strongly-ordered counterterms, starting from the 
expression of the real-virtual counterterm $K_{n+1}^{(\mathbf{RV})}$, such that the combination 
$K^{\RV}_\npo + I^{\otwo}_\npo$ is free of IR poles. For the sake of illustration, we focus on 
the soft component. We begin by constructing the cross-section-level radiative soft function 
from factorisation, using \eq{eq:Sng1}. We find
\beq
\label{sigmalevS}
  S_{n,g}^{(1)} \big( \{\beta_i\}; k \big) & = &
  {\s_{n,g}^{(0)}}^{\dagger} \big( \{\beta_i\}; k \big)
  \left[
  S_{n+1}^{(1)} \big( \{\beta_i\}, \beta_k \big) - J_{\E_g}^{(1)} (\beta_k)
  \right]
  \s_{n,g}^{(0)} \big( \{\beta_i\}; k \big)
  \nnb\\ &&
  + \, \int\frac{d^d\ell}{(2\pi)^d} \,
  \Big( \s_{n,g}^{(0) \mu} \big( \{\beta_i\}; \ell \big) \Big)^{\dagger} \,
  J_{g,g}^{(1) \mu \nu} (\ell; k) \,\,
  \s_{n,g}^{(0)\nu} \big( \{\beta_i\}; \ell \big) \, .
\eeq
We now use the finiteness conditions, suitably replacing one-loop factors with tree-level 
radiative factors. This leads to
\beq
\label{eq:RV12cancelSoft}
  S_{n,g}^{(1)} \big( \{\beta_i\}; k_1 \big)  \! & + & \!
  \int d \Phi(k_2) \,
  \bigg\{ S_{n;g,g}^{(0)} \big( \{\beta_i\}; k_{[12]}; k_2 \big) \\ 
  & - &
  \Big( \! \hspace{1pt} \mathcal{S}_{n,g}^{(0)} \big( \{\beta_i\}; k_{[12]} \big) 
  \! \Big)^{\! \dagger} J_{\E_g,g}^{(0)} (\beta_{k_1}; k_2) \, 
  \mathcal{S}_{n,g}^{(0)} \big( \{\beta_i\}; k_{[12]} \big)
  \nn \\ & + & \!
  \int \frac{d^d\ell}{(2\pi)^d} \, 
  \Big( \! \hspace{1pt} \mathcal{S}^{(0) \, \mu }_{n,g} \big( \{\beta_i\}; \ell \big) 
  \! \Big)^{\! \dagger}
  \! \sum_{f_1, f_2} J_{g, f_1f_2}^{(0) \, \mu \nu} (\ell; k_1, k_2) \, \,
  \mathcal{S}_{n,g}^{(0) \, \nu} \big( \{\beta_i\}; \ell \big)
  \bigg\} \, = \, \text{finite} \, .
  \nn 
\eeq
We see that the refactorisation of strongly-ordered soft radiation suggests an expression for 
the soft component of the strongly-ordered counterterm $K_{n+2}^{(\mathbf{12})}$. Indeed, we can take the integrand appearing in \eq{eq:RV12cancelSoft} as a definition of the local 
counterterm. With simple steps, one gets
\beq
\label{eq:K12s_2}
  K_{n+2}^{(\mathbf{12}, \, \text{s})}
  \, = \,
  {\mathcal{H}_n^{(0)}}^{\dagger} 
  \sum_{f_1,f_2}
  \bigg[
  S^{(0)}_{n;f_{[12]},f_2} \big( k_{[12]}, k_2 \big)
  +
  S^{(0)}_{n,f_{[12]}}
  \Big(
  J_{f_{[12]},f_1f_2}^{(0)}
  -
  J_{\E_{[12]},f_2}^{(0)}
  \Big)
  \bigg] \,
  \mathcal{H}_{n}^{(0)}
  \, .
\eeq
In the soft sector, $K_{n+2}^{(\mathbf{12}, \, \text{s})}$ cancels all poles of 
$K_{n+1}^{(\mathbf{RV}, \, \text{s})} $ by construction. In fact, the explicit poles of the 
soft component $K_{n+1}^{(\mathbf{RV})}$ are encoded in the radiative, one-loop soft function, 
so that we have
\beq
  K_{n+1}^{(\mathbf{RV}, \, \text{s})} \, = \,  
  {\mathcal{H}_n^{(0)}}^{\dagger} \, S_{n,g}^{(1)} \, \mathcal{H}_n^{(0)} \, + \, \text{finite} \, .
\eeq
It is then straightforward to verify the cancellation occurring between $K_{n+1}^{(\mathbf{RV}, 
\, \text{s})}$ and $K_{n+2}^{(\mathbf{12}, \, \text{s})}$, upon integrating the latter over 
$d \Phi(k_2)$. The same steps apply to the collinear case, where the hard-collinear component 
of the real-virtual counterterm can be written as 
\beq
\label{KRVhci1}
  K^{(\mathbf{RV}, \, \text{hc})}_{n+1, \,  i}
  & = &  
  {\hard_n^{(0)}}^{\dagger}
  \sum_{f_1, f_2}
  J_{f_i, f_1 f_2}^{(0), \, \text{hc}} \,
  \Big[
  S_3^{(1)} - J_{\E_i}^{(1)} +
  \sum_{k=1}^2 J_{f_k, f_k}^{(1), \, \text{hc}}
  \Big] \,
  {\cal H}^{(0)}_n
  \, ,
\eeq
and the corresponding strongly-ordered counterterm reads
\begin{eqnarray}
\label{eq:K12hci}
  K_{n+2,i}^{(\mathbf{12}, \, \text{hc})}
  & = &
  {\hard_n^{(0)}}^{\dagger} \!\!\!
  \sum_{f_1,f_2,f_3}
  \bigg[
  J_{f_i,f_1f_2}^{(0), \, \text{hc}}(\bar k_1,\bar k_2) \, S_{3,f_3}^{(0)} -
  J_{f_i,f_1f_2}^{(0), \, \text{hc}}(k_1,k_2) \, J_{\E_i,f_3}^{(0)}
  \nnb \\
  &&
  \qquad\qquad
  + \!\!\!
  \sum_{kl=\{12,21\}}
  J_{f_i,f_{[k3]} f_l}^{(0), \, \text{hc}} \,
  \Big(
  J_{f_{[k3]},f_kf_3}^{(0)} - J_{\E_k,f_3}^{(0)}
  \Big)
  \bigg] \,
  \hard_n^{(0)}
  \, .
\end{eqnarray}
Again, the pole cancellation between \eq{KRVhci1} and the integral of \eq{eq:K12hci} can be 
easily proven by exploiting finiteness relations involving one-loop and radiative soft and jet 
functions. 


\section{Outlook}
\label{Outlo}

In this contribution we have outlined a procedure to identify local subtraction counterterms 
describing the singular behaviour of real-radiation in soft and collinear limits, paying 
special attention to strongly-ordered configurations. We have presented the general structure 
of subtraction counterterms at NNLO, and we have sketched it at N$^3$LO, displaying a pattern  
that facilitates the generalisation to higher orders. Our procedure for constructing counterterms
starts from the all-order factorisation of virtual amplitudes, and uses finiteness relations 
to deduce the form of infrared counterterms for real radiation, reversing the standard approach 
to infrared subtraction\footnote{A similar perspective has been recently explored by other groups 
in the context of \emph{antenna subtraction}~\cite{Braun-White:2023sgd} and \emph{nested 
soft-collinear subtraction}~\cite{Devoto:2023rpv}.}. All counterterms are expressed in terms of 
gauge-invariant matrix elements of fields and Wilson lines. We have focused on the main 
challenge associated with strongly-ordered configurations, namely  ensuring that the integrals of the 
corresponding counterterms cancel the poles of mixed real-virtual contributions. Even if our 
approach does not immediately translate into a concrete subtraction algorithm, it provides 
crucial insights on the architecture of infrared subtraction to all orders. Further work is underway to build an algorithmic implementation of these ideas.


\section*{Acknowledgments}
\label{Ackno}

\noindent We are grateful to Sandro Uccirati for participating in the initial stages of this 
project. PT has been supported by the Italian Ministry of University and Research (MUR), grant 
PRIN 2022BCXSW9, and by Compagnia di San Paolo, grant TORP S1921 EX-POST 21 01.


\end{document}